\documentclass[12pt, preprint]{aastex}
\usepackage{graphicx}

\shorttitle{Korreck et. al. } \shortauthors{SN1006 Non Radiative
Collisionless Shock-FUSE Data}

\begin{document}

\title{FUSE Observation of the Non-Radiative Collisionless
Shock in the Remnant of SN1006}
\author{K. E. Korreck\altaffilmark{1,2}, J.C. Raymond\altaffilmark{2}, T.H. Zurbuchen\altaffilmark{1}, P. Ghavamian\altaffilmark{3}}

\altaffiltext{1}{University of Michigan, Department Of
Atmospheric, Oceanic and Space Sciences, 2455 Hayward, Ann Arbor MI
48109} \altaffiltext{2}{Harvard-Smithsonian Center for
Astrophysics,
    60 Garden Street, Cambridge, MA 02138}
\altaffiltext{3}{Department of Physics and Astronomy, Johns
Hopkins University, 3400 North Charles Street, Baltimore, MD,
21218-2686 }

\begin{abstract}
The appearance of the young supernova remnant SN1006 is
dominated by emission from non-radiative shocks in the NE
and NW regions.  At X-ray energies the northeast shock
exhibits predominantly nonthermal synchrotron emission,
while the northwest shock exhibits a thermal spectrum.  We
present far ultraviolet spectra of the northeast (NE)  and
northwest (NW) portions of SN 1006 acquired with the Far
Ultraviolet Spectroscopic Explorer (FUSE).  We have
detected emission lines of \ion{O}{6} (1032, 1038 \AA) and
broad Ly-$\beta$ (1025 \AA) in the NW filament, but detect
no emission lines in the NE region down to a level of 4.7
$\times$ 10$^{-17}$ erg cm$^{-2}$ s$^{-1}$
arcsecond$^{-2}$.  We observed in the NW an
\ion{O}{6} intensity of 2.0 $\pm$ 0.2 $\times$ 10$^{-16}$
erg cm$^{-2}$ s$^{-1}$ arcsecond$^{-2}$ and measured an
\ion{O}{6} line width of 2100 $\pm$ 200 km s$^{-1}$ at a
position where the H$\alpha$ width was measured to be 2290
$\pm$ 80 km s$^{-1}$ (Ghavamian et al.  2002).  This
implies less than mass proportional heating of the ions.  
Using the ratio of intensities,
I(NW)/I(NE)$\sim$n(NW)/n(NE), the density ratio of the two
regions is found to be $\ge$ 4, a value that is with the
uncertainties of the ratio of 2.5 measured by Long et al.
(2003).  The derived \ion{O}{6} kinetic temperature is
compared to previous estimates of electron, proton, and ion
temperatures in the remnant to study the relative heating
efficiency of various species at the shock front.  The
degree of postshock temperature equilibration may be
crucial to particle acceleration since the temperature of
each species determines the number of high speed particles
available for injection into an acceleration process that
could produce Galactic cosmic rays.
\end{abstract}



\section{Introduction}

SN1006 (G327.6+14.6) is a nearby Type $\textrm{I}$a supernova remnant at a distance of
2.1 kpc \citep{wink03}.  With a mean expansion rate of 8700 km s$^{-1}$ it is $\sim$18
pc wide \citep{wink03}. The remnant has a high Galactic latitude and modest foreground
reddening, E(B-V)=0.11 $\pm$ 0.02 \citep{sch80}.  This young supernova remnant is
entering the Sedov-Taylor phase of supernova remnant evolution.

SN1006 has been observed at radio \citep{pye81}, optical
\citep{gha02,kwc87,smi91}, ultraviolet \citep{ray95} and X-ray
\citep{wink03,lon03,bam03} wavelengths. Gamma ray observations
\citep{tan98} were reported but not confirmed.  Thin, pure Balmer
line filaments were found in the optical. In the radio and X-ray,
the remnant has a limb-brightened shell structure with cylindrical
symmetry around a SE to NW axis probably aligned with the ambient
galactic magnetic field \citep{rey86,jp88}.  The NE shock front of
SN1006 shows strong non-thermal X-ray and possible gamma ray
emission while the NW shock shows very little non-thermal emission
at radio or X-ray wavelengths.

Ly-$\beta$, \ion{He}{2}, \ion{C}{4}, and \ion{O}{6} lines were observed
from the faint optical Balmer line filament of the NW shock of the
supernova remnant, by the Hopkins Ultraviolet Telescope (HUT), flown
during the Astro-2 space shuttle mission.  The observed FWHM of the lines
were 2230, 2558, 2641 km s$^{-1}$, respectively (the \ion{O}{6} line width
could not be measured).  A kinetic temperature could be calculated from
these line widths.  The kinetic temperatures of these species are not
equal, because they do not scale inversely with the square root of their
atomic mass. Instead, the UV observations do suggest that
$\rm{T}_{\rm{ion}} \sim \frac{\rm{m}_{\rm{ion}}}{\rm{m}_{\rm{p}}}
\rm{T}_{\rm{proton}}$ indicating little to no temperature equilibration
between species.

SN1006 provides an opportunity to investigate parameters of
non-radiative collisionless shocks faster than 2000 km s$^{-1}$.
Collisionless shocks appear in many astrophysical phenomena, from
coronal mass ejections (CMEs) in the heliosphere to jets in
Herbig-Haro objects.  When a shock is non-radiative the detection
of emission from the shock front is possible, as all of the
optical and UV emission of a non-radiative shock comes from a
narrow zone directly behind the shock front. Interactions at the
collisionless shock front depend upon mechanisms such as plasma
waves to transfer heat, kinetic energy and momentum, and it is not
well understood how particles of different masses and charges are
affected by these processes. The temperature of the species and
the degree of temperature equilibration between electrons, protons
and other ions is central to the interpretation of X-ray spectra,
which effectively measure electron temperature. The energy
distribution of a particle species is important to cosmic ray
studies as only those particles at a high energy tail of a
particle distribution are available for cosmic ray acceleration.

The method of using H$\alpha$ lines to determine collisionless
shock parameters was originated by \citet{cr78} and \citet{kcr}.
The H$\alpha$ line has a two component profile. The width of the
broad component of the H$\alpha$ line is related to the post-shock
proton temperature as a result of charge exchange between neutrals
and protons, which produces a hot neutral population behind the
shock.  The narrow component of the H$\alpha$ line is produced
when cold ambient neutrals pass through the shock and emit line
radiation before being ionized by a proton or electron.  The ratio
of the broad to narrow flux is sensitive to electron-ion
equilibrium and the pre-shock neutral fraction. The FWHM of
H$\alpha$ line was measured to be 2290 $\pm$ 80 km s$^{-1}$, with
models implying the speed of the shock is v$_{\rm{shock}}= 2890
\pm 100$ km s$^{\rm{-1}}$ (Ghavamian et. al. 2002; hereafter
GWRL02). The H$\alpha$ broad to narrow intensity ratio measured to
be 0.84 implies an electron temperature much lower than the ion
temperature.

This UV observation from the FUSE satellite focused on the shock
front in the NW observed by HUT and GWRL02 and on a region in the NE dominated by non-thermal
emission.  From the spectra, a broad Lyman $\beta$ line
($\lambda$$\sim$1025 \AA) and the doublet of \ion{O}{6}
($\lambda$$\sim$1032, 1038 \AA) were analyzed for spectral width,
intensity, and flux.  We use the line widths of the NW and the
intensities of the \ion{O}{6} lines in the NE and NW shock fronts
to compare the electron-ion and ion-ion temperature equilibration
efficiencies as well as densities.  The heating of different
particle species by the shock front as well as parameters of
collisionless shocks that affect particle species heating will be
discussed.

\section{Observations}

The Far Ultraviolet Spectroscopic Explorer (FUSE) has a wavelength
range of approximately 900-1180 \AA.  The Large Square Aperture
(LWRS), with a field-of-view of 30\arcsec\ x 30\arcsec, with a
roll angle of 167$^{\rm{o}}$, was chosen for this observation
because models predicted that the \ion{O}{6} emission behind the
shock would be spread over 35\arcsec\ \citep{ray95,lam96}.  The
LWRS has a filled-aperture resolution of about 100 km s$^{-1}$.
The FUSE spectrometer consists of four independent channels with
two segments each. Four of these eight segments operate in the
wavelength range for the \ion{O}{6} doublet, $\lambda$=1031.91,
1037.61 \AA.   However, the Silicon Carbon (SiC) coated channels,
because they are optimized for $\lambda$ $\le$ 1020 \AA, add an
unacceptable amount of noise to our faint signal, so only the
Lithium Floride (LiF) channels are used.  These two segments are
designated LiF1A and LiF2B.  The LiF1A channel covers wavelengths
987.1 - 1082.3 \AA, while the LiF2B covers 979.2-1075.0 \AA.

 Although the northwest region of
the remnant has been observed before in the UV \citep{ray95}, we
have much better spectral resolution and a more optimal aperture
size to include the entire ionization region given that it may be
larger than 19\arcsec  \citep{lam96}. The apertures used for past
observations were HUT=19\arcsec\ x 197\arcsec\ \citep{ray95} and
CTIO RC Spectrometer=2\arcsec x 51\arcsec\ \citep{wink03}(GWRL02).
We positioned the aperture center to be 5\arcsec-10\arcsec\ behind
the H$\alpha$ filament where the peak formation of \ion{O}{6}
occurs. The NE position was chosen based on the edge of the X-ray
filament from \citet{lon03}.

FUSE observations of the northwest region, centered at
$\alpha_{2000}$ =$15^{\rm{h}}$ $2^{\rm{m}}$ $19.17^{\rm{s}}$,
$\delta_{2000}$ =-41$^o$ 44\arcmin\ 50.4\arcsec\, were obtained on
23 June 2001 and 26 February 2002 with total exposure times of
35,627 s and 6,690 s. Observations of the northeast region,
centered at $\alpha_{2000}$=$15^{\rm{h}}$ $4^{\rm{m}}$
$5.0^{\rm{s}}$, $\delta_{2000}$=-41$^o$ 50\arcmin\ 40.5\arcsec\,
were obtained on 25 June 2001 and 27 February 2002 with exposure
times of 42,365 s and 9,666 s. The locations of observations are
shown superimposed on an H$\alpha$ image of the remnant taken with
the CTIO Schmidt telescope in Figure \ref{halpha} \citep{wink03}.
Inserted in the figure is a close up from Chandra \citep{lon03} of
the NE region of observation to illustrate the x-ray morphology,
although no optical emission is obviously present.

There are four components to the background of this observation;
detector background, geocoronal lines, the diffuse galactic UV
continuum and diffuse galactic \ion{O}{6} emission.   The
background count distribution on the FUSE detectors is composed of
two separate components \citep{fuse03}.  The `intrinsic'
background forms from the $\beta$-decay of potassium in the
microchannel plate (MCP) detector glass and the spacecraft
radiation environment. The effect of the spacecraft radiation
environment on the detector background varies from night to day
and with solar activity, but over a short observing time this
variation is not significant.  The second component is caused by
scattered light, primarily geocoronal Ly-$\alpha$. This line
produces detector averaged count rates as small as 20$\%$ of the
intrinsic background during the night and increasing to 1-3 times
the intrinsic rate during the day.  The other two components of
the background, galactic UV emission and diffuse \ion{O}{6}
emission, will be discussed later.

The observations were calibrated with the CalFUSE Pipeline Version
2.2.1. Data from all exposures are processed through the pipeline
and then co-added following the FUSE Data Analysis Cookbook and
The FUSE Observer's Guide. The data were selected to contain only
the night observations. This greatly reduces the geocoronal
background. The night-only exposure times were 32,287 s for the
Northwest and 39,386 s for the Northeast.

\section{Analysis and Results}

As mentioned above, the background consists of detector noise, geocoronal lines,
diffuse galactic
\ion{O}{6} and an
astrophysical UV continuum. The first two sources were explained
in the previous section, but the additional diffuse UV continuum
must be treated separately. It does not originate from SN1006, as
it is seen in both of the entirely different regions of the remnant; the NE shock and 
the NW
shock. The diffuse background is attributed to light from hot
stars scattering on dust.  The diffuse UV continuum is especially
bright in this region of the sky according to models by
\citet{mur95}.  A value of $8.4\times 10^{-15}$ erg cm$^{-2}$
s$^{-1}$ \AA$^{-1}$ was quoted by \citet{ray95} while we are
seeing approximately $6.5\times 10^{-15}$ erg cm$^{-2}$ s$^{-1}$
\AA$^{-1}$ through an aperture one quarter the size of the HUT observation.

In addition to the diffuse UV continuum, \citet{she01,she02} and \citet{ott03} have
found a
diffuse \ion{O}{6} background.  The brightness of the \ion{O}{6} background is 4700
$\pm$ 2400 photons cm$^{-2}$ s$^{-1}$ sr$^{-1}$\citep{ott04}.
The widths of the diffuse \ion{O}{6} lines fall between 10 and 160 km s$^{-1}$.  In
the current NE spectrum diffuse \ion{O}{6} emission has a width of $\le$ 200 km
s$^{-1}$ and a brightness of 3500 photon cm$^{-2}$ s$^{-1}$ sr$^{-1}$.  We attribute
the NE emission to the diffuse galactic \ion{O}{6} background. This
enabled us to subtract the
NE as a background from the NW data to further eliminate airglow lines, the diffuse UV
emission and the galactic \ion{O}{6} background. The intensities of the airglow lines
at 1042\AA\ and 1048\AA\ are quite similar in both the NE and NW, further allowing
this subtraction.  The raw spectra of the NW and the NE regions are shown in Figure
\ref{fullspectra}, with airglow lines marked.

For the NW region, a nonlinear chi-squared minimization routine
was used to fit Gaussian line profiles to the spectra. The
wavelengths considered for analysis were restricted to 1010-1050
\AA\ to minimize spurious background effects near the ends of the
detector's spectral range. The data were binned by 0.1 \AA\ to
increase the number of counts per bin without losing resolution,
as the line widths were several Angstroms wide. The width of the
broad H$\alpha$ line, from GWRL02, is v$_{\rm{H}}$=2290 km
s$^{-1}$. Since the Ly-$\beta$ line is formed by the same
process\citep{kcr}, its line width was set equal to the H$\alpha$
broad component width . The shift of the centroid of the broad and
narrow component of H$\alpha$, v=29 km s$^{-1}$, is effectively
negligible (implying that the shock is viewed completely edge-on)
so the broad Ly-$\beta$ line centroid was fixed at its rest
wavelength. Only the intensity of the line was a free parameter.
The blue wing of the line was fit from 1010 \AA\ to 1024.5 \AA.
Due to the extinction from interstellar dust, a correction factor must be applied
to deredden the observed flux.  Using the extinction curves of \citet{ccm89}, the
resulting dereddened Ly-$\beta$ flux is 2.3 $\pm$ 0.3 x
10$^{\rm{-13}}$ erg cm$^{-2}$ s$^{-1}$.

After subtracting the
fitted broad Ly-$\beta$ line profile, the wavelength range from
1022-1028 \AA\ was excluded from the fitting routine in order to
avoid negative fluxes and residual airglow that would skew the
gaussian fits of the \ion{O}{6} lines. At $\sim$1037.0 \AA\ there
were absorption features present that coincided with a \ion{C}{2}
line and molecular hydrogen lines, along with an \ion{O}{1}
airglow line. The absorption feature with the spectral range from
1035 - 1038 \AA\, was therefore excluded from the fit.

The \ion{O}{6} doublet was fit with two gaussians with fixed
centers at 1031.91 and 1037.61 \AA\ respectively corresponding to
the centroid of H$\alpha$. The doublet line intensities were
forced to have a 2:1 ratio but the magnitude of the intensities
were allowed to vary. The observed flux is 6.7 $\pm$ 0.1 x
10$^{-17}$ erg cm$^{-2}$ s$^{-1}$ arcsec$^{-2}$. Total dereddened
flux for the \ion{O}{6} doublet lines was 1.8 $\pm$ 0.2 x
10$^{-13}$ erg cm$^{-2}$ s$^{-1}$. The \ion{O}{6} line widths were
measured to be 7.2 $\pm$ 0.4 \AA\ FWHM, or equivalently 2100 $\pm$
100 km s$^{-1}$. The formal error on the fit is 100 km s$^{-1}$.
However, due to systematic error a more conservative error of
$\pm$ 200 km s$^{-1}$ is used.  The fits are shown in Figure 3.
The width is within the limiting estimate of \citet{ray95} of
$\le$ 3100 km s$^{-1}$ and is within 1$\sigma$ error of the
H$\alpha$ width of 2290 km s$^{-1}$. Although the faint signal in
the NE did not allow for a statistically significant fit, an upper
limit of \ion{O}{6} intensity was found assuming a width of 2000
km s$^{-1}$. The observed upper limit on the \ion{O}{6} line is
1.6 x 10$^{-17}$ erg cm$^{-2}$ s$^{-1}$ arcsecond$^{-2}$.  An
upper limit on the dereddened intensity of \ion{O}{6} in the NE is
4.2 $\times$ 10$^{-14}$ erg cm$^{-2}$ s$^{-1}$.  The upper limit
of flux for Ly-$\beta$ in the NE is 1.6 x 10$^{-17}$ erg cm$^{-2}$
s$^{-1}$ arcsecond$^{-2}$. An upper limit to the dereddened
Ly-$\beta$ intensity in the NE region is 4.4 $\times$ 10$^{-14}$
erg cm$^{-2}$ s$^{-1}$.

Past observations of SN1006 line widths and intensities are
summarized in Table 1.  In order to compare past measurements made
with varying aperture sizes, we use intensity per arcsecond
measured along the length of the filament.  The Ly-$\beta$ from
the HUT and the current FUSE observation are consistent.  We can
use the various measurements to study the ion heating.  The proton
temperature was found using the shock speed of 2890 km s$^{-1}$
from GWRL02.  This proton temperature was then multiplied by
m$_{ion}$/m$_{p}$ to calculate the mass proportional temperatures.
These calculated temperatures were then compared to the
temperatures given by using the FWHM of each ion line.  The
temperature of \ion{O}{6} as indicated by its FWHM is less than
mass proportional by 48\%.  For the other ions, \ion{He}{2}, \ion
{C}{4} the heating was also less than mass proportional, by 21\%
and 18\% respectively.

The brightness of the \ion{O}{6} lines is proportional to density,
n$_{0}$, and the depth of the filament along the line of sight.
Therefore, an upper limit to the density in the NE can be found by
the ratio of intensities provided that the depths along the line
of sight are known. \citet{lon03} calculated a density ratio of
n(NW)/n(NE) = 2.5.  From the thermal component of the Chandra
X-ray spectra \citet{lon03} estimated a pre-shock ISM density of
0.25 cm$^{-3}$ in the NW. Using the limit to the \ion{O}{6}
intensity ratio of the NW and NE a ratio of the densities is found
to be n(NW)/n(NE) $\ge$ 4, which is within the uncertainties of the Long et al. 
calculations. Therefore, assuming a pre-shock density
in the ISM of 0.25 cm$^{-3}$ in the NW, the pre-shock NE density
$\le$ 0.06 cm$^{-3}$.  This density calculation depends on the
assumptions of similar depths along the line of sight in the NE
and the NW and of similar numbers of \ion{O}{6} photons per atom
passing through the shock.  The amount of electron-ion
equilibration in the NE would affect these assumptions.  Greater
electron-ion equilibration in the NE would increase the number of
\ion{O}{6} photons per atom \citep{lam96}, so the limit on the
density in the NE would be even smaller.  We attribute the low
upper limit on the \ion{O}{6} intensity in the NE to the low
density medium into which the remnant is expanding.

\section{Discussion}

\ion{O}{6} lines were not conclusively observed in the faint non-radiative
non-thermal NE shock indicating that the two distinct shock regions heat ions 
differently.  Ion heating is important to cosmic ray acceleration and the 
overall energy distribution of the system.  The ions have most of their kinetic energy 
in a broad distribution which is generally non-Maxwellian as the time to equilibrium 
via Coulomb collisions for ions and protons is 1.2 $\times$ 10$^{5}$ years.  To 
understand the heating at the shock front, turbulence, line widths, methods of 
calculating heating, and the role of neutrals at the shock front will be discussed.

\subsection{Small Scale Turbulence}
Turbulence plays a role in the evolution of fast shocks in 
supernova remnants\citep{rey04,ell92}.  Small scale 
turbulence spreads the line profile of an ion much like thermal broadening of a line 
profile.  Since some of the shock energy must be used for 
bulk flow, we will examine turbulence with a velocity of 1500 km s$^{-1}$ which is 
large enough to affect the spectra but not contain all the energy of the flow.  
Turbulence decays on a time scale proportional to the characteristic length of the 
turbulence divided by the velocity of the turbulence $\sim$ $\ell$/v \citep{ten57}.  
The width of the H$\alpha$ filament is at most 10$^{16}$ cm based on its 1\arcsec\ 
apparent 
width on the sky(GWRL02), making the time scale of the 
turbulence 10$^{8}$ s $\sim$ 3 years.  Using this decay time and the post-shock speed of 750 km 
s$^{-1}$, one 
quarter of the shock speed, the post-shock region affected by turbulence would be 7.5 $\times$ 10$^{15}$ cm.  The 
\ion{O}{6} 
filament  
with an observed width of 3 $\times$ 10$^{17}$ cm, assuming the 
30\arcsec\  FUSE 
aperture is filled, is also too wide to be dominated by turbulence thus the turbulence 
that 
is present in the shock of SN1006 is short-lived and not a major 
source of line broadening.   

\subsection{Line Widths of \ion{O}{6}, UV lines and H$\alpha$}

The UV line profiles of the current observations can be compared
with past observations of various ion species. The currently
observed \ion{O}{6} line width in the NW shock is within 1$\sigma$
of the H$\alpha$ line width previously measured by GWRL02.
\citet{vink03} measured an \ion{O}{7} line width of 3.4 $\pm$ 0.5
eV, or approximately 1775 $\pm$ 261 km s$^{-1}$ from a different
northwest region. This line width is substantially narrower than
those of other ion species measured thus far, although the region
of observation for this measurement is different from the position
of our observations. Along a 124\arcsec\ slit, \citet{smi91} found
little variation in the H$\alpha$ profiles, indicating that the
oxygen temperature does not vary significantly along the length of the NW filament. 
 This implies one
of two processes. First, the line width could decline with
ionization state and distance behind the shock due to Coulomb
collisions, as Coulomb collisions would transfer heat to other
species.  In Section 4.1, we found the Coulomb collision time to
be far too long for this process to be important.  The second more
probable scenario is that some of the lower temperature \ion{O}{7}
is from the reverse shock in the supernova ejecta. The detection
of \ion{Si}{13} and \ion{Mg}{11} X-ray lines \citep{lon03} in the
NW region of the remnant agrees with the hypothesis that the
emission is coming from ejecta near the shock front.

To find a proton temperature we have thus far used the width of
the H$\alpha$ line.  However, the proton thermal speed is not
simply equal to the velocity derived from the width of the
H$\alpha$ line at high temperatures.  The cross section for
neutral-proton charge transfer, the process that produces the
broad H$\alpha$, falls off at high energies allowing for the neutral
hydrogen distribution function to be narrower than that of the
protons\citep{kcr}. This results in an H$\alpha$ profile that
would incorrectly indicate a lower temperature than the
actual proton temperature.

\subsection{Heating at the Shock Front}
From the current observations there are two ways to calculate the
temperature of the ions.  The first method to calculate the temperature is
based
on the thermalization of the bulk velocity of the shock.  The
second method uses the FWHM of the gaussian line fits as the thermal
velocity that can be used to find the temperature.  To determine
the kinetic temperature of a species from the bulk thermalization we use
the equation,
\begin{equation}
kT_{i}=\frac{3}{16}m_{i}v_{\rm{shock}}^{2}
\end{equation}

\noindent where the subscript $i$ indicates the species, $k$ is the Boltzman
constant, $T$ is temperature, $m_{i}$ is the mass of the species
and $v_{\rm{shock}}$ is the shock speed = 2890 km s$^{-1}$(GWRL02). This
gives a temperature for \ion{O}{6} of 2.9 $\times$ 10$^{9}$ K and
for the protons of 1.8 $\times$ 10$^{8}$ K.  The ratio of the temperatures
is mass proportional, T$_{\rm{oxygen}}$=16T$_{\rm{proton}}$, which is
expected using this method. This
heating occurs when some fraction of the energy of the shock speed
is transferred to the thermal velocity of the protons or ions.

Using the width of the \ion{O}{6} lines to calculate temperature,
the temperature is measured to be 1.5 $\times$ 10$^{9}$ K.  The
\ion{O}{6} temperature derived from the observed line width is
less than that predicted by the kinetic temperature equation for
no equilibration among particle species.  The ratio of the
temperatures indicates that \ion{O}{6} is heated to a temperature 48\% less than the 
value predicted for mass proportional heating.  Ions are being
heated by a process other than the bulk fluid velocity
thermalization or there is a heat loss mechanism for the ions.

Heating of ions in collisionless shocks has been studied by
\citet{ber97} using heliospheric shock data.  In examming
\ion{O}{7}, it was found that the oxygen was preferentially heated
19-39 times more than the protons.  In studying the solar wind,
\citet{lee00} assume greater than mass-proportional heating as
part of the coronal heating process.  As a consequence, ions
non-adiabatically expand upstream (not being reflected by the
shock front) and move with a velocity equal to their gyration
velocity as they go upstream. These hot highly energized ions
could act as a precursor that takes away a significant amount of
energy.

The current supernova observation of less than mass proportional
heating lies in stark contrast to the heliospheric collisionless
shocks.  Several factors and processes determine the extent of ion
heating.  The first comparison to be made is the speed of the
shock relative to the local Alfv\'{e}nic speed.  The solar shocks
propagate at 400-1000 km s$^{-1}$.  SN1006's shock is propagating
at almost 3000 km s$^{-1}$.  The Alfv\'{e}nic Mach number, the
ratio of the shock speed to the square root average of the thermal and local 
Alfv\'{e}nic speed, is $\le$
10 for solar shocks but upwards of 200 for the supernova shock.
The orientation of the magnetic field with respect to the normal
of the shock front is also of importance as quasi-perpendicular
shocks and quasi-parallel shocks are quite different.  If the
current magnetic field orientation for SN1006 is correct, the NW is propagating 
parallel to the ambient magnetic field, while both parallel and perpendicular shocks 
are observed in the solar wind.

A measure of the importance of the magnetic field is the parameter
$\beta$.  The plasma $\beta$, the ratio of thermal to
magnetic pressure upstream, is small ($\le 1$) for heliospheric shocks.
Using
the parameters for SN1006 and the general value for the Galactic plane ISM magnetic 
field
($\sim$3 $\mu$G), the NE has $\beta=0.02$ and in the NW $\beta =
0.1$.  This indicates that the magnetic field pressure dominates
over the thermal pressure at the ISM/remnant boundary as in the solar 
wind, in contrast 
to the 
ISM which is assumed to have a $\beta$ of unity.  It is possible that the
change in density from pre-shock to
post-shock conditions is an important characteristic in the propagation
and heating of ions in the collisionless shock fronts.

In order to
determine the cause of the different heating found in the heliosphere and
the supernova further investigation of the influence of pressure,
density, Mach number and velocity on ion heating by the shock is
necessary.

\subsection{Neutrals at the Shock Front}
In the analysis scheme used here from \citet{cr78} and
\citet{kcr}, neutrals play a vital role.  Neutrals undergo charge
exchange or emit line radiation to produce the H$\alpha$ and
Ly-$\beta$ emission.  There are fast neutrals produced by the
shock, as evident by the broad components of the H$\alpha$ and
Ly-$\beta$ lines. This could create a neutral precursor for the
shock\citep{smi94, lim96}. The hydrogen and oxygen neutral
fractions are tightly coupled by charge transfer, thus information
about the neutral fraction of hydrogen can be used to diagnose the
neutral fraction of oxygen. These neutrals should become pickup
ions like those seen in the solar wind \citep{val76} when they
pass through the shock and become ionized.  Pickup ions like those
in the heliosphere can then act as a high energy seed population
for Fermi acceleration just as heliospheric pickup ions are the
seed population for anomalous cosmic rays\citep{fis74}.

The \ion{He}{2} 4686\AA\ line can be used as an indicator of
neutral fraction due to its insensitivity to pre-shock neutral
fraction and electron-ion pre-shock temperature equilibrium
(GWRL02). In the NW, observations have shown \ion{He}{2} emission
lines \citep{ray95}(GWRL02).  The ratio of \ion{He}{1}/\ion{He}{2}
can then be used to find an H neutral fraction which is a
parameter in the relation of the H$\alpha$ two component intensity
ratio, I$_{\rm{broad}}$/I$_{\rm{narrow}}$, and the electron-ion
temperature ratio.  In addition, GWRL02 calculated the pre-shock H
population to be 90\% ionized but the pre-shock He population is
70\% neutral. Using the H$\alpha$ broad-to-narrow intensity ratio
calculated for 90\% pre-ionized medium, the temperature ratio,
T$_{\rm{electron}}$/T$_{\rm{proton}}$, was found to be $\le$ 0.07,
showing little to no equilibration between protons and electrons.
Using the ratio of T$_{\rm{electron}}$/T$_{\rm{proton}}$, we find
an electron temperature of $\le$ 1.2 $\times$ 10$^{7}$ K,
approximately 1 keV, which is an upper limit that agrees with the
value found by \citet{lon03} of T$_{\rm{electron}}$ $\le$ 0.6 keV,
but significantly less than the oxygen and proton temperatures
found for this observation(1.5 $\times$ 10$^{9}$ K and 1.8
$\times$ 10$^{8}$ K, respectively).

\section{Summary}
In summary, the two shock regions of SN1006 studied here provide a unique cosmic 
laboratory for shocks and their acceleration processes.  Clearly, the properties of 
the interstellar medium play a crucial role in shaping these shocks.  We conclude with 
the following summary of our observations and interpretations.

\begin{enumerate}
\item The material that the NE shock front is encountering is less dense than in the NW region, with a ratio n(NW)/n(NE)$\ge$4, and is best seen in the
X-ray or radio wavelengths.  The NW shock front could be moving
into a diffuse \ion{H}{1} cloud or similarly dense region.

\item The \ion{O}{6} line width of the NW shock indicates that oxygen ions are 
heated to a temperature less than 48\% of the value predicted by mass proportional 
heating.  This differs from the
observations of other non-radiative collisionless shock fronts
such as those in the heliosphere which found ion temperatures 20-40 times in excess of 
the values predicted by mass proportional heating.  The roles of density, pressure, 
magnetic
field orientation with respect to the shock normal, velocity and
Mach number should be examined to better determine the ion heating
mechanisms.

\item The plasma at the shock front has not had time to come to equilibrium via
Coulomb collisions.  The plasma is in a non-equilibrium state with
energy distributed differently between species of the plasma. This
is in agreement with the work on temperature equilibrium done by
\citet{gha02} who found the ratio of proton to electron
temperature to be $\le$ 0.07 indicating a plasma far from equilibrium.
\end{enumerate}

The role of the neutral fraction of the ISM population in the charge exchange
interaction should be
examined in detail as it may greatly affect the outcome of the
shock-ISM interaction.  The rate at which the plasma becomes
isotropic, the particle distribution, and the time
scale to reach isotropic and Maxwellian conditions are in need of examination to
understand the heating process present in the collisionless shock.
Further work will be done to model the plasma conditions in
collisionless shock fronts to include neutral fraction as well as
examine the role of electron population on \ion{O}{6} formation.  This
work should help advance the understanding of the shock acceleration of
particles
and the physics of a collisionless shocks in varying environments.

\acknowledgments

This work is based on observations made with the NASA-CNES-CSA Far
Ultraviolet Spectroscopic Explorer. FUSE is operated for NASA by the Johns Hopkins
University under NASA contract NAS5-32985.  We gratefully acknowledge conversations
and help with FUSE night only data extractions by R. Sankrit. This work is supported
by NASA Grant number NEG5-10352 given to the Smithsonian Observatory. K. Korreck is
supported by the University of Michigan NSF-Rackham Engineering Award. This work made
use of the NASA Astrophysics Data System (ADS).

\clearpage



\begin{figure}
\plotone{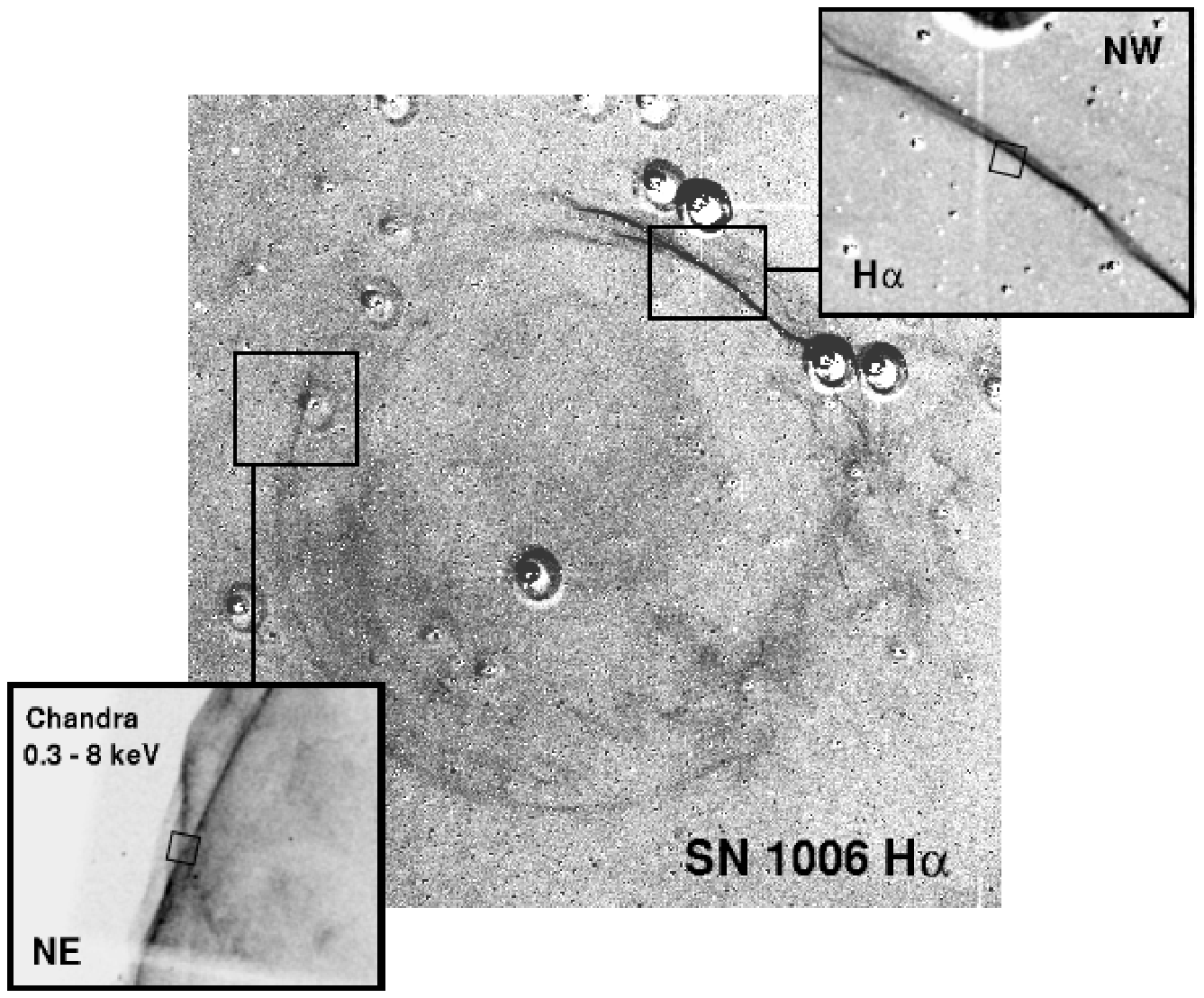}
\end{figure}

\begin{figure}
\plotone{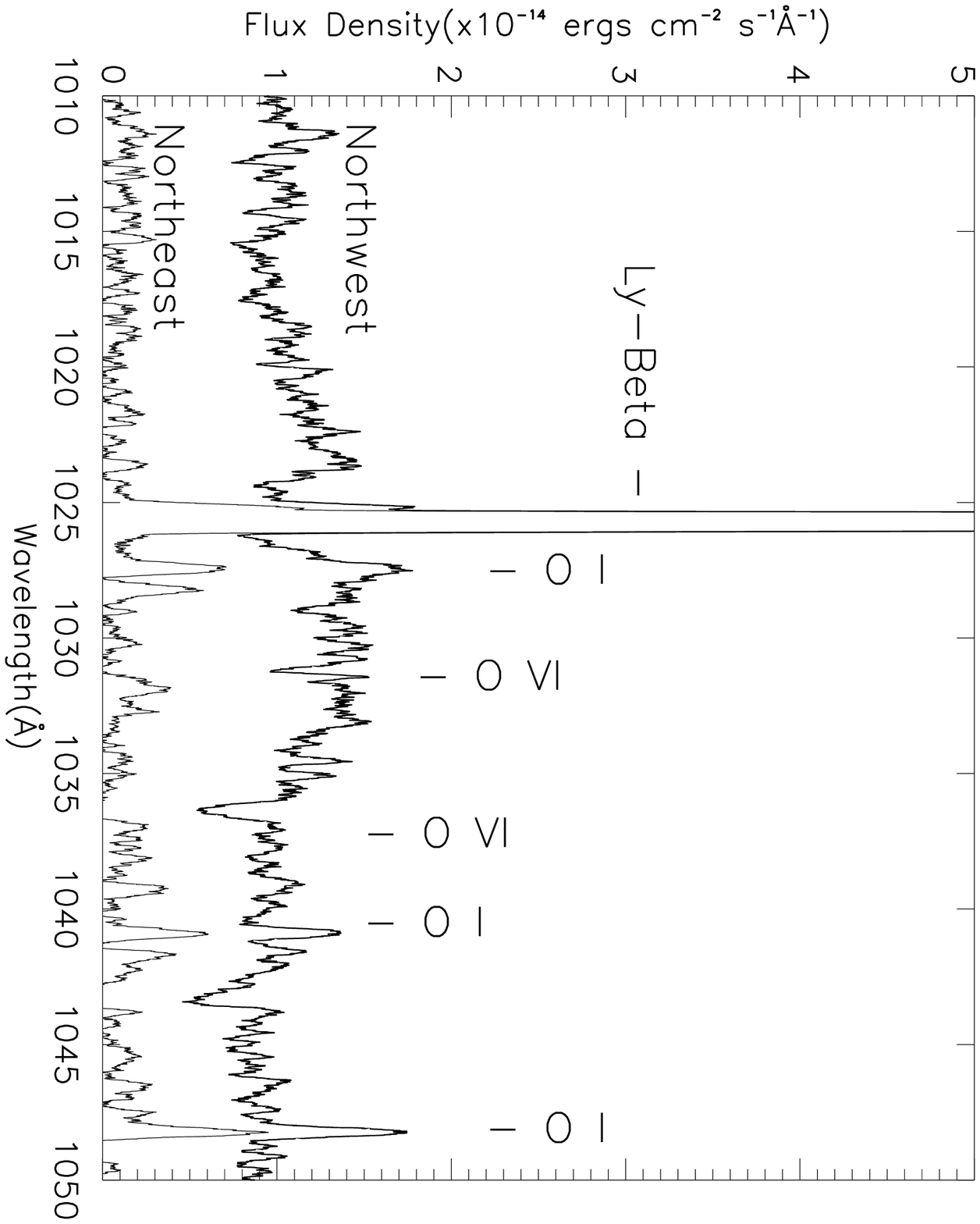}
\end{figure}

\begin{figure}
\plotone{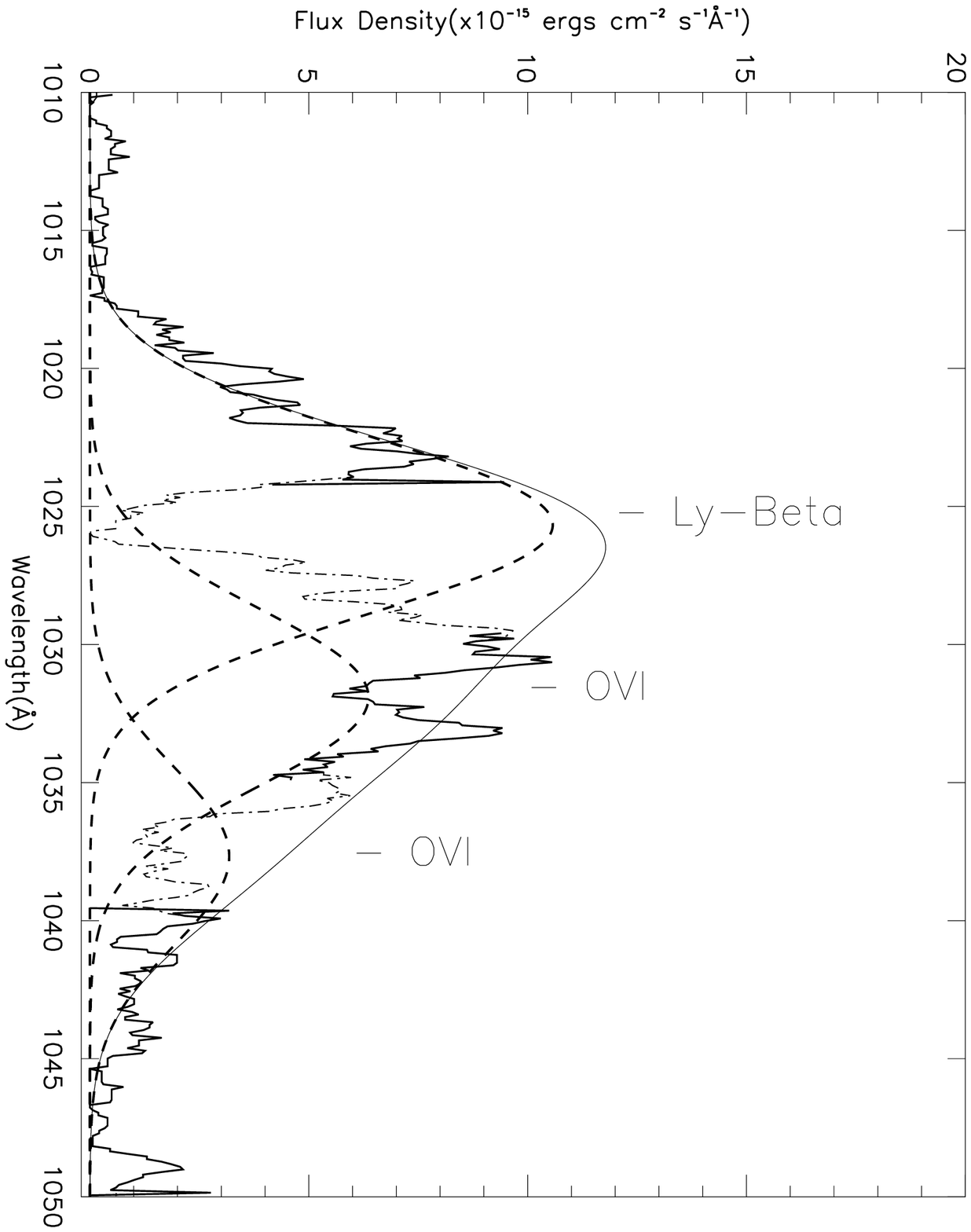}
\end{figure}

\clearpage

\figcaption[f1.eps]{H$\alpha$ image of SN1006 taken by the CTIO Schmidt telescope from 
Winkler
et al. (2003). Closeup images of the observed filaments are shown in the
insets.
In each case the interior box, drawn to scale, shows the location of
the 30\arcsec\ x 30\arcsec\  FUSE LWRS aperture.  The NW blowup is
from the same H$\alpha$ image while the NE blowup is a 0.3-8 keV Chandra
image (Long et al. 2003).
\label{halpha}}

\figcaption[f2.eps]{Raw FUSE spectra from the Northeast
and Northwest region of SN1006.  The NW is offset from the NE by
0.5 for clarity. Geocoronal lines are marked.  The Ly-$\beta$ peak dominates in both
spectra.
The NE is consistently fainter than the NW, but the geocoronal
line intensities are similar. \label{fullspectra}}

\figcaption[f3.eps]{FUSE spectra from the NW, binned at 0.1
\AA\, with the NE subtracted as background.  The dotted dashed lines are the
regions of the spectra that were excluded from the fits.  The dashed lines are
the fits for the Ly-$\beta$, \ion{O}{6} 1032 and 1037 \AA\  lines,
with FWHM of 2290 and 2100 km s$^{-1}$ respectively. The solid line represents the
addition of the fits of the three spectral lines.\label{fit1}}
\clearpage


\begin{deluxetable}{ccccccc}
\tabletypesize{\scriptsize}
\tablecolumns{7} \tablewidth{0pc} \tablecaption{Summary of UV
Emission Lines in NW Filament of SN1006}

\tablehead{ \colhead{Ion} & \colhead{Intensity} & \colhead{Filament Length} &
\colhead{FWHM}&
\colhead{Temperature} &
\colhead{m$_{ion}$/m$_{p}$T } & \colhead{\%} \\
&(photons cm$^{-2}$s$^{-1}$arcsec$^{-1}$)&
(arcsec) & (km
s$^{-1}$) &(Kelvin)&(Kelvin) &Mass\\
&(x10$^{-4})$ & &Observed&from FWHM && Prop\\

}
\startdata

  H-$\alpha$$^{1}$ &2.1 &51 & 2290 $\pm$ 80&1.8 $\times$ 10$^{8}$ \tablenotemark{a} &
-
& - \\ \hline

  Ly-$\beta$ &4.0 &30 & 2290(fixed)& -  & -  & - \\ \hline

He II$^{2}$ &0.99 &197 & 2558$\pm$618& 5.7 $\times$ 10$^{8}$ & 7.2
$\times$ 10$^{8}$  &
79\% \\\hline

  C IV$^{2}$ &1.7& 197& 2641$\pm$ 355& 1.8$\times$ 10$^{9}$
  &2.2$\times$ 10$^{9}$
 &82\% \\\hline

  O VI &3.1&30 &2100 $\pm$ 200&1.5$\times$ 10$^{9}$
 &2.9$\times$ 10$^{9}$
 &52\% \\ \hline
  O VII$^{3}$& & 60 &1775$\pm$261& 1.1 $\times$ 10$^{9}$
&2.9$\times$ 10$^{9}$ & 38\% \\\hline

\enddata
\tablenotetext{a}{Temperature derived from shock speed of 2890 km
s$^{-1}$.} \tablerefs{(1) Ghavamian et al. 2002; (2) Raymond et
al. 1995; (3) Vink et al. 2003}

\end{deluxetable}




\begin{thebibliography}{}
\bibitem[Anderson, Sankrit, Dupuis(2003)]{fuse03}Anderson, B. G.,
Sankrit, R., Dupuis, J., 2003, FUSE Observer's Guide
\bibitem[Bamba et al.(2003)]{bam03} Bamba, A., Yamazaki, R., Ueno, M., Koyama, K.,
2003, \apj, 589, 827
\bibitem[Berdichevsky et al.(1997)]{ber97} Berdichevsky, D., Geiss,
J., Gloeckler, G., Mall, U., 1997, \jgr, 102, 263
\bibitem[Cardelli, Clayton, Mathis(1989)]{ccm89}Cardelli, J. A.,
Clayton, G. C., Mathis, J. S., 1989, \apj, 345, 245
\bibitem[Chevalier, Kirshner, \& Raymond(1980)]{kcr} Chevalier,
R. A., Kirshner, R. P., Raymond, J. C., 1980, \apj, 235, 186
\bibitem[Chevalier \& Raymond(1978)]{cr78}Chevalier, R. A.,
Raymond, J. C., 1978, \apj, 225, L27
\bibitem[Dyer et al.(2004)]{dye04} Dyer, K. K., Reynolds, S. P.,
Borkowski, K. J., 2004, \apj, 600, 752
\bibitem[Ellison \& Reynolds(1991)]{ell92} Ellison, D.C., Reynolds, S. P., 1991, \apj, 
382, 242
\bibitem[Fisk, Kozlovsky \& Ramaty(1974)]{fis74}Fisk, L. A., Kozlovsky, B., Ramaty, R.
1974, \aplett, 190, L35
\bibitem[Ghavamian et al.(2002)]{gha02}Ghavamian, P., Winkler, P. F., Raymond, J. C., Long, K. S. 2002, \apj, 572, 999
\bibitem[Hester, Raymond, and Blair(1994)]{hes94}Hester, J. J., Raymond, J. C., Blair,
W. P., 1994, \apj, 420, 721
\bibitem[Henry \& Murthy(1993)]{hm93} Henry, R. C., Murthy,
J., 1993, \apj, 418, L17
\bibitem[Jones \& Pye(1988)]{jp88} Jones, L. R., Pye, J. P., 1989, \mnras, 238, 567
\bibitem[Kirshner, Winkler, and Chevalier(1987)]{kwc87}Kirshner, R. P.,
Winkler, P. F., Chevalier, R.A., 1987, \apj, 315, L135
\bibitem[Laming et al.(1996)]{lam96}Laming, J. M., Raymond, J.
C., McLaughlin, B. M., Blair, W. P., 1996, \apj, 472, 267
\bibitem[Lee \& Wu(2000)]{lee00}Lee, L. C., Wu, B. H., 2000, \apj, 535,
1014
\bibitem[Lim \& Raga (1996)]{lim96}Lim, A. J., Raga, A. C., 1996, \mnras, 280, 103
\bibitem[Long et al.(2003)]{lon03} Long, K. S., Reynolds, S. P.,
Raymond, J. C., et.al. 2003, \apj, 586, 1162
\bibitem[Mancuso et al.(2002)]{man02} Mancuso, S., Raymond, J. C.,
Kohl, J., et. al. 2002, \aap, 383, 267
\bibitem[Murthy(2002)]{mur02} Murthy, J., 2002, \aj, 23, 23
\bibitem[Murthy \& Henry(1995)]{mur95} Murthy, J., Henry, R. C.,
1995, \apj, 448, 848
\bibitem[Otte et. al. (2003)]{ott03}Otte, B., Dixon, W. V., Sankrit, R., 2003, \apj,
586, L53
\bibitem[Otte, Dixon, \& Sankrit (preprint)]{ott04}Otte, B., Dixon, W. V., Sankrit,
R., 2004, \apj, preprint
\bibitem[Pye et al.(1981)]{pye81}Pye, J. P., Pounds, K. A., Rolf,
D. P., Smith, A., Willingate, R., Seward, F. D., 1981, \mnras, 194,
569
\bibitem[Raymond, Blair, Long(1995)]{ray95}Raymond, J. C., Blair, W. P., Long,
K. S., 1995, \apj, 454, L31
\bibitem[Reynolds(2004)]{rey04}Reynolds, S. P., 2004, Adv. Space Res., 33, 461
\bibitem[Reynolds(1996)]{rey96}Reynolds, S. P., 1996, \apj, 459, L13
\bibitem[Reynolds \& Chevalier(1981)]{rey81}Reynolds, S. P., Chevalier, R. A., 1981,
\apj, 245, 912
\bibitem[Reynolds \& Gilmore(1986)]{rey86}Reynolds, S. P., Gilmore, D. M.,
1986, \aj, 92, 1138
\bibitem[Schweizer \& Middleditch(1980)]{sch80}Schweizer, F.,
Middleditch, J., 1980, \apj, 241, 1039
\bibitem[Shelton et al.(2002)]{she02}Shelton, R. L., 2002, \apj, 569, 758
\bibitem[Shelton et al.(2001)]{she01}Shelton, R. L., Kruk, J. W., Murphy, E. M., et
al., 2001, \apj, 560, 730
\bibitem[Smith et al.(1991)]{smi91}Smith, R. C., Kirshner, R. P.,
Blair, W. P., Winkler, P. F., 1991, \apj, 375, 652
\bibitem[Smith et al.(1994)]{smi94}Smith, R. C., Raymond, J. C., Laming, J. M., 1994,
\apj, 420, 286
\bibitem[Spitzer (1956)]{spi56}Spitzer Jr., L., 1956, 'Physics of Fully
Ionized Gases', Interscience
Publishers, Inc.
\bibitem[Tennekes \& Lumley (1972)]{ten57}Tennekes, H., Lumley, J.L., 1972, 'A First 
Course in Turbulence', The MIT Press
\bibitem[Tanimori(1998)]{tan98}Tanimori, T., 1998, \iaucirc, 188, 121
\bibitem[Vasyliunas \& Siscoe (1976)]{val76}Vasyliunas, V. M., Siscoe, G. L., 1976,
\jgr, 81, 1247
\bibitem[Vink et al.(2003)]{vink03}Vink, J., Laming, J. M., Gu, M. F.,
Rasmussen, A., Kaastra, J. S., 2003, \apj, 587, L31
\bibitem[Willingdale et al.(1996)]{wil96}Willingdale, R., West, R. G.,
Pye, J. P., Stewart, G. C., 1996, \mnras, 278, 749
\bibitem[Winkler, Gupta, Long(2003)]{wink03}Winkler, P. F., Gupta, G.,
Long, K. S., 2003, \apj, 585, 324
\end{thebibliography}
\end{document}